# The swap operation in the two-qubit Heisenberg XXZ model ——effects of anisotropy and magnetic field

Yue Zhou[1], Guo-Feng Zhang[2*], Fu-Hua Yang[1], and Song-Lin Feng[1]

[1]*State Key Laboratory for Superlattices and Microstructures, Institute of Semiconductors, Chinese Academy of Sciences, P. O. Box 912, Beijing 100083, People's Republic of China*

[2] *Department of Physics, School of Sciences, Beijing University of Aeronautics & Astronautics, Xueyuan Road No. 37, Beijing 100083, People's Republic of China*

**Abstract:** In this paper we study the swap operation in a two-qubit anisotropic XXZ model in the presence of an inhomogeneous magnetic field. We establish the range of anisotropic parameter $\lambda$ within which the swap operation is feasible. The swap errors caused by the inhomogeneous field are evaluated.

## I. INTRODUCTION

Quantum computers (QCs) have attracted much attention recently due to their potential to greatly exceed their classical counterparts [1-4]. Of the various schemes that have been proposed, the ones based on solid state systems are believed to have the best scalability. Moreover, the solid state schemes can largely take advantage of modern semiconductor technology and micro-fabrication technology. Because of these advantages, the study of solid state quantum computation is currently an important field [5–9].

One of the primary conditions for realizing quantum computation is to represent quantum information robustly. The spins of electrons or nuclei live in the Hilbert space spanned by the spin-up and spin-down states, so that they are natural candidates to represent qubits. Another fundamental condition required for quantum computation is the universal quantum gates that implement the unitary transformations [10]. It has been proved that any multi-qubit unitary transformations can be implemented by a combination of single-qubit gates and CNOT gates, so they are universal. In spin-based proposals single-qubit gates correspond to single-spin rotations, which can be achieved by a magnetic field pulse in the *x-y* plain or g-factor engineering [11]. The implementation of a two-qubit gate is based mainly on the exchange interaction between spins. Among two-qubit gates, the swap gate $U_{SW}$ plays a fundamental role. The swap gate itself is not universal; but it has been proved that its square root $U_{SW}^{1/2}$ is universal. The CNOT gate can be constructed by combining of single-qubit operations and $U_{SW}^{1/2}$ [5], so that along with the single-qubit gates any unitary transformation can be achieved.

In most proposed QCs, the energy difference between the qubit states is large compared to the qubit-qubit interaction, which corresponds to a system of spins in a strong external magnetic field [8]. In solid state structures, inhomogeneity of the field is inevitable, so it's necessary to consider the effect of an inhomogeneous field. In Ref. [12] Hu et al studied the effect of an inhomogeneous magnetic field in an XXX-type (i.e., $J_x = J_y = J_z$) coupling system. However, in real systems, the spin-spin interaction may be anisotropic due to surface and interface effects, as well as spin-orbit coupling [13]. In order to investigate the influence of the anisotropy, more general models should be considered, for instance, the XXZ model (i.e., $J_x = J_y \neq J_z$). Moreover, in Ref. [12], the effect of an inhomogeneous field is only demonstrated in a simple situation, which is insufficient and needs further proof. These are the main motivations of this paper.

---

[*] Corresponding author; electronic mail address: gf1978zhang@163.com

## II. THE MODEL AND METHOD

Under an inhomogeneous magnetic field, the effective Hamiltonian of a two-qubit Heisenberg XXZ model is given by [14]:

$$H = \frac{1}{2}\left[ J(\sigma_1^x \sigma_2^x + \sigma_1^y \sigma_2^y + \lambda \sigma_1^z \sigma_2^z) + (B+b)\sigma_1^z + (B-b)\sigma_2^z \right], \tag{1}$$

where $B \geq 0$ is the magnetic field along the $z$ direction. The parameter $b$ represents the degree of inhomogeneity. $J$ is the exchange coupling in the $x$-$y$ plane. The chain is called antiferromagnetic if $J > 0$ and ferromagnetic if $J < 0$. The parameter $\lambda$ is the anisotropy in the $z$ direction. For $\lambda = 1$ we call it an XXX chain; for $\lambda = 0$ we call it an XX chain. In other words, the XXX and XX models can be regarded as special cases of the XXZ model. Note that we are working in such units that $B$, $b$ and $J$ are dimensionless.

We define $|1\rangle$ and $|0\rangle$ as the spin-up and spin-down states, respectively. The eigenstates and corresponding eigenvalues of the Hamiltonian (1) can be expressed as

$$\psi_1 = |00\rangle, \qquad E_1 = \frac{1}{2}(\lambda J - 2B),$$

$$\psi_2 = |11\rangle, \qquad E_2 = \frac{1}{2}(\lambda J + 2B),$$

$$\psi_3 = \frac{\varepsilon}{\sqrt{J^2 + \varepsilon^2}}|10\rangle + \frac{J}{\sqrt{J^2 + \varepsilon^2}}|01\rangle, \qquad E_3 = -\frac{\lambda J}{2} - \eta,$$

$$\psi_3 = \frac{\zeta}{\sqrt{J^2 + \zeta^2}}|10\rangle + \frac{J}{\sqrt{J^2 + \zeta^2}}|01\rangle, \qquad E_4 = -\frac{\lambda J}{2} + \eta, \tag{2}$$

where $\eta = \sqrt{b^2 + J^2}$, $\varepsilon = b - \eta$, and $\zeta = b + \eta$.

Assuming that the initial state is given by $\varphi(0) = \varphi_1(1,0) \otimes \varphi_2(2,0)$, the first and second positions in the bracket indicate the sequence number of the spins and the time $t$, respectively. It then evolves under the Hamiltonian (1):

$$\varphi(t) = e^{-iHt}\varphi(0). \tag{3}$$

If the wave function becomes $\varphi = \varphi_2(1,0) \otimes \varphi_1(2,0)$ at some time, then the swap operation is achieved. Due to the exchange interaction of the spins, the two spins will generally be entangled. Only at certain times will the system evolve into disentangled states. To find whether the swap operation is feasible, we can check whether the state corresponds to a swapped one when it is disentangled. Suppose the initial state is:

$$\varphi(0) = (\alpha_1 |1\rangle + \alpha_2 |0\rangle) \otimes (\beta_1 |1\rangle + \beta_2 |0\rangle). \tag{4}$$

We expand the initial state in the basis of eigenstates of the Hamiltonian (1), and Eq. (3) becomes

$$\varphi(t) = a_1(t)\psi_1 + a_2(t)\psi_2 + a_3(t)\psi_3 + a_4(t)\psi_4, \tag{5}$$

where $a_1 = \alpha_2 \beta_2 e^{-iE_1 t}$,

$$a_2 = \alpha_1 \beta_1 e^{-iE_2 t},$$

$$a_3 = \frac{(\alpha_2 \beta_1 \zeta - \alpha_1 \beta_2 J)\sqrt{J^2 + \varepsilon^2}}{2\eta J} e^{-iE_3 t} \equiv \frac{M\sqrt{J^2 + \varepsilon^2}}{2\eta J} e^{-iE_3 t},$$

$$a_4 = \frac{(\alpha_1 \beta_2 J - \alpha_2 \beta_1 \varepsilon)\sqrt{J^2 + \zeta^2}}{2\eta J} e^{-iE_4 t} \equiv \frac{N\sqrt{J^2 + \zeta^2}}{2\eta J} e^{-iE_4 t}.$$

If a two-qubit system is in a disentangled state, the reduced density matrix of either spin is pure. The reduced density matrix of the first spin is given by

$$\rho_{1\uparrow\uparrow} = |b_2|^2 + |b_3|^2, \qquad \rho_{1\downarrow\downarrow} = |b_1|^2 + |b_4|^2,$$

$$\rho_{1\uparrow\downarrow} = b_1^* b_3 + b_2 b_4^*, \qquad \rho_{1\downarrow\uparrow} = b_1 b_3^* + b_2^* b_4, \tag{6}$$

with $b_1 = \alpha_2 \beta_2 e^{-iE_1 t}$, $\qquad b_2 = \alpha_1 \beta_1 e^{-iE_2 t}$,

$$b_3 = \frac{\varepsilon M e^{-iE_3 t} + \zeta N e^{-iE_4 t}}{2\eta J}, \qquad b_4 = \frac{M e^{-iE_3 t} + N e^{-iE_4 t}}{2\eta}.$$

For a pure state density matrix $\rho = |\mu\rangle\langle\mu|$, in two and higher dimensional Hilbert space, there is at least one state $|v\rangle$ that is orthogonal to $|\mu\rangle$. Then we have $\rho|v\rangle = |\mu\rangle\langle\mu|v\rangle = 0$, which means that the set of homogeneous linear equations $\rho|v\rangle = 0$ has nonvanishing solutions, so the coefficient matrix $\rho$ satisfies

$$\det \rho = 0. \tag{7}$$

It can be proved that in two-dimensional Hilbert space, Eq. (7) is also the sufficient condition for the density matrix to be pure [15]. Hence, the state of the first spin is pure if and only if $\rho_{1\uparrow\uparrow}\rho_{1\downarrow\downarrow} - |\rho_{1\uparrow\downarrow}|^2 = 0$. From Eq. (6) we have

$$\rho_{1\uparrow\uparrow}\rho_{1\downarrow\downarrow} - |\rho_{1\uparrow\downarrow}|^2$$
$$= |b_1 b_2 - b_3 b_4|^2$$

$$= \left| \frac{\left[Jb(\alpha_2^2\beta_1^2 - \alpha_1^2\beta_2^2) - 2J^2\alpha_1\alpha_2\beta_1\beta_2\right][1-\cos(2\eta t)] - iJ\eta(\alpha_2^2\beta_1^2 + \alpha_1^2\beta_2^2)\sin(2\eta t)}{2\eta^2} + \right.$$

$$\left. \alpha_1\alpha_2\beta_1\beta_2(1-e^{-i2\lambda Jt}) \right|^2. \tag{8}$$

## III. INFLUENCE OF THE ANISOTROPY INTERACTION

First we look at the homogeneous field situation. Here we presume $J > 0$, and it can be proved that the result is the same for $J < 0$. When $b \to 0$, we have $\varepsilon \to -J$, $\zeta \to J$, $\eta \to J$, then Eq. (8) becomes

$$\rho_{1\uparrow\uparrow}\rho_{1\downarrow\downarrow} - |\rho_{1\uparrow\downarrow}|^2$$

$$= \left| \alpha_1\alpha_2\beta_1\beta_2[\cos(2Jt) - \cos(2\lambda Jt)] - i[\frac{1}{2}(\alpha_2^2\beta_1^2 + \alpha_1^2\beta_2^2)\sin(2Jt) - \alpha_1\alpha_2\beta_1\beta_2\sin(2\lambda Jt)] \right|^2. \tag{9}$$

Now consider the value of $t$ for which Eq. (9) vanishes. If $t$ depends on the initial state parameters $\alpha_1$, $\alpha_2$, $\beta_1$ and $\beta_2$, then for an unknown initial state the swap operation can't be realized; so $t$ must be independent of the initial state parameters. Thus in order to obtain $\frac{1}{2}(\alpha_2^2\beta_1^2 + \alpha_1^2\beta_2^2)\sin(2Jt) - \alpha_1\alpha_2\beta_1\beta_2\sin(2\lambda Jt) = 0$, we must have $\sin(2Jt) = \sin(2\lambda Jt) = 0$. When the above condition is satisfied, we have $\cos(2Jt) = \pm 1$ and $\cos(2\lambda Jt) = \pm 1$. For the XXX model, $\lambda = 1$, so $\cos(2Jt) - \cos(2\lambda Jt) = 0$ holds naturally; for the XXZ model where $\lambda \neq 1$, however, $\cos(2Jt) = \cos(2\lambda Jt)$ is also necessary, so we can see that $t$ is tied to the value of $\lambda$.

Case 1: If $\lambda$ is a rational number, then it can always be written in the form: $|\lambda| = \frac{m}{n}$, where $m$ and $n$ are co-prime integers.

① For $\cos(2Jt) = \cos(2\lambda Jt) = 1$, the solution is $t = \frac{kn}{J}\pi$, where $k = 1, 2, 3...$ is the number of the period.

② For $\cos(2Jt) = \cos(2\lambda Jt) = -1$, if both $m$ and $n$ are odd, the solution is $t = \frac{(2k-1)n\pi}{2J}$; if either $m$ or $n$ is even, however, $\cos(2Jt) = -1$ and $\cos(2\lambda Jt) = -1$ can't be achieved simultaneously, then there is no $t$ independent of $\alpha$ and $\beta$ for which $\rho_{1\uparrow\uparrow}\rho_{1\downarrow\downarrow} - |\rho_{1\uparrow\downarrow}|^2 = 0$.

Case 2: If $\lambda$ is an irrational number, $\cos(2Jt) = \cos(2\lambda Jt) = \pm 1$ can't be strictly achieved. But for an appropriate $t$ it may be approximately achieved, so that the wave function is close to a pure state.

Now let's focus on the state of the first spin while $\rho_{1\uparrow\uparrow}\rho_{1\downarrow\downarrow} - |\rho_{1\uparrow\downarrow}|^2 = 0$. When $\cos(2Jt) = \cos(2\lambda Jt) = 1$, $\varphi_1 = \frac{1}{\sqrt{2}}(\alpha_1|1\rangle + (-1)^{k(m+n)}e^{iBt}\alpha_2|0\rangle)$, the state returns to the initial one except for an additional phase shift. When $\cos(2Jt) = \cos(2\lambda Jt) = -1$, $\varphi_1 = \frac{1}{\sqrt{2}}(\beta_1|1\rangle + (-1)^{\frac{n-m}{2}}e^{iBt}\beta_2|0\rangle)$, the states of the two spins are swapped except for an additional phase shift, so that the swap operation is achieved after the additional phase shift is corrected by a single-spin operation.

In order to correct the additional phase shift, it's necessary to determine its expression. In the XXX model, the additional phase shift is always $e^{iBt}$, whether for 'return' or 'swap' operations. In the XXZ model, however, the phase shift is also related to $\lambda$ and $k$: For the 'return' operation, if $m+n$ is even, the phase shift is $e^{iBt}$, which is the same as the XXX model; if $m+n$ is odd, then when $k$ is even or odd, the phase shift is $e^{iBt}$ or $e^{-Bt}$, respectively. For swap operation, when $\frac{n-m}{2}$ is even or odd, the additional phase shift is $e^{iBt}$ or $e^{-iBt}$, respectively. The reason is that for some values of $\lambda$, while $\cos(2Jt) = \cos(2\lambda Jt) = \pm 1$, we have $\cos(Jt) = -\cos(\lambda Jt)$ or $\sin(Jt) = -\sin(\lambda Jt)$, so a phase shift of $e^{i\pi}$ is induced.

Finally, we consider the swap operation while either $m$ or $n$ is even. As is shown above, in this case an ideal swap operation is not feasible. However, if $\cos(2\eta t) \approx \cos(2\lambda\eta t) \approx -1$, the swap operation can be achieved approximately. The criterion can be roughly measured by $\tau(\lambda) = \min[2 + \cos(2\eta t) + \cos(2\lambda\eta t)]$, the smaller $\tau$ is, the closer the swap operation is to an ideal one. It can be proved that $\tau$ decreases monotonously with the increase of both $m$ and $n$. In other words, if both $m$ and $n$ are small (e.g. $\frac{1}{2}$, $\frac{2}{3}$, $\frac{1}{4}$ ...), even an approximate swap operation is irrealizable. Another important issue is the duration that the operation takes. If $\lambda$ is in the neighborhood of above values, though theoretically feasible, it'll take rather a long time to realize the swap operation, which means that the operation time before decoherence is quite limited. On the contrary, the operation duration is relatively short when $\lambda$ is close to the value that both $m$ and $n$ are small odd numbers.

## IV. INFLUENCE OF AN INHOMOGENEOUS FIELD

We now turn to the influence of an inhomogeneous magnetic field. First we look at a simple situation, in which $\beta_1 \cdot \beta_2 = 0$. Without loss of generality, let $\beta_1 = 1$ and $\beta_2 = 0$, so that $\varphi(0) = (\alpha_1|1\rangle + \alpha_2|0\rangle)|1\rangle$. It can be calculated that if and only if $\cos(2Jt) = \cos(2\lambda Jt) = -1$ the result is closest to an ideal swap operation. Defining $b = \delta J$, we have

$$\rho_{1\uparrow\uparrow} = 1 - \frac{|\alpha_2|^2 \delta^2}{1+\delta^2} \tag{10}$$

It can be seen from Eq. (10) that the first spin is definitely upward only when $|\alpha_2| = 0$ (i.e., the initial state is the eigenstate of the Hamiltonian and the system is in a stationary state). For other values of $|\alpha_2|$, there would be an error of magnitude $\Delta = \frac{|\alpha_2|^2 \delta^2}{1+\delta^2} \approx |\alpha_2|^2 \delta^2$. Considering the constraint of contemporary available error correction schemes, the swap operation is feasible only when $|\delta| \ll 1$, or rather when the order of $b$ is far lower than $J$. Now let us look at more general situations. When $\cos(2Jt) = \cos(2\lambda Jt) = -1$, if $\beta_1 \cdot \beta_2 \neq 0$, the reduced density matrix is given by

$$\rho_{1\uparrow\uparrow} = |\beta_1|^2 + \frac{(|\alpha_1\beta_2|^2 - |\alpha_2\beta_1|^2)\delta^2 + (\alpha_1^*\alpha_2\beta_1\beta_2^* + \alpha_1\alpha_2^*\beta_1^*\beta_2)\delta}{1+\delta^2} \equiv |\beta_1|^2 + C$$

$$\rho_{1\downarrow\downarrow} = |\beta_2|^2 - C$$

$$\rho_{1\uparrow\downarrow} = (-1)^{\frac{n-m}{2}} \beta_1\beta_2^* \frac{\beta_1\beta_2^* + \alpha_1\alpha_2^*(|\beta_2|^2 - |\beta_1|^2)\delta}{\beta_1\beta_2^*\sqrt{1+\delta^2}} e^{-iBt} \equiv (-1)^{\frac{n-m}{2}} \beta_1\beta_2^* D e^{-iBt} \tag{11}$$

There are infinitely many ways to express a mixed state as a mixture of nonorthogonal pure states; however, all the different ways make no difference for experimental measurements [16]. Hence, only one kind of expression is needed. Here we write it as:

$$\varphi(t) = \begin{cases} p_1 : \beta_1|1\rangle + (-1)^{\frac{n-m}{2}} \beta_2 e^{i(Bt-\theta)}|0\rangle \\ p_2 : |1\rangle \\ p_3 : |0\rangle \end{cases} \tag{12}$$

where $p_1 = |D|$, $p_2 = (1-|D|)|\beta_1|^2 + C$, $p_3 = (1-|D|)|\beta_2|^2 - C$, $\theta = \arg D$.

Whether the swap operation is well achieved can be measured by the parameter $p$, which is the probability that we measure the state $\varphi(t)$ in the basis $\beta_1|1\rangle + (-1)^{\frac{n-m}{2}} \beta_2 e^{iBt}|0\rangle$. For the ideal situation, $p = 1$; and the closer $p$ is to 1, the better the swap operation is achieved. For the state expressed by Eq. (12), we have

$$p = p_1 \left| |\beta_1|^2 + |\beta_2|^2 e^{-i\theta} \right|^2 + p_2 |\beta_1|^2 + p_3 |\beta_2|^2 \tag{13}$$

Inserting the expressions of $p_1$, $p_2$ and $p_3$ into Eq. (13), we obtain the error magnitude of swap operation:

$$\Delta = 1 - p \approx (|\beta_1|^4 - 2|\alpha_1 \beta_1|^2 + |\alpha_1|^2)\delta^2 \tag{14}$$

In the proof, we use a power series expansion and neglect the third- and higher-order terms. Note that

$$|\beta_1|^4 - 2|\alpha_1 \beta_1|^2 + |\alpha_1|^2 \geq |\beta_1|^4 - 2|\alpha_1 \beta_1|^2 + |\alpha_1|^4 = (|\beta_1|^2 - |\alpha_1|^2)^2 > 0$$

$$|\beta_1|^4 - 2|\alpha_1 \beta_1|^2 + |\alpha_1|^2 < \max(|\alpha_1|^2, 1 - |\alpha_1|^2) \leq 1 \tag{15}$$

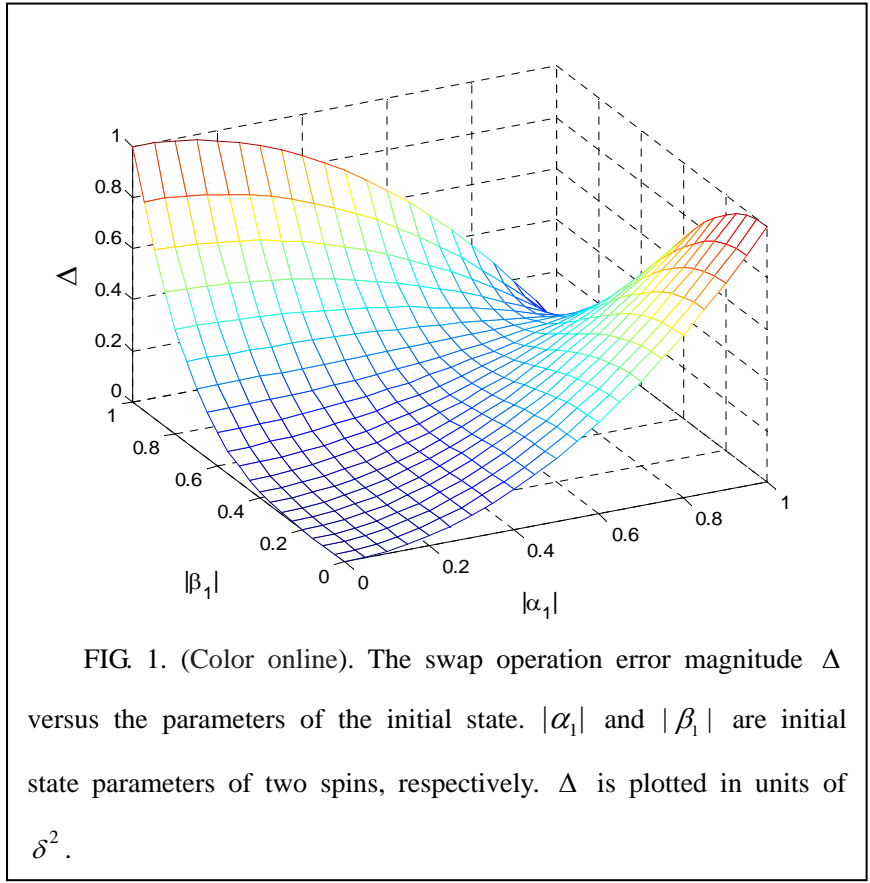

FIG. 1. (Color online). The swap operation error magnitude $\Delta$ versus the parameters of the initial state. $|\alpha_1|$ and $|\beta_1|$ are initial state parameters of two spins, respectively. $\Delta$ is plotted in units of $\delta^2$.

Combining this result with the situation in which $\beta_1 \cdot \beta_2 = 0$, we have $0 \leq \Delta \leq \delta^2$, which is demonstrated in Fig. 1. In other words, for any initial state, the error magnitude of the swap operation caused by the inhomogeneous magnetic field is bounded by $\delta^2$. Note that the phase shift caused by the inhomogeneous magnetic field doesn't qualitatively affect the operation, which can also be elicited by the qualitative analysis below: If $|\beta_1| \approx |\beta_2|$, then $\theta$ is so trivial that can be ignored; if $|\beta_1| \cdot |\beta_2| \approx 0$, the wave function itself is close to either of the standard bases $|1\rangle$ and $|0\rangle$, so that the phase difference between them is of little significance.

## V. CONCLUSION

We have investigated the influence of the anisotropy in the spin-spin interaction. Whether the swap operation is still feasible largely depends on the form of the anisotropy parameter $\lambda$. We also prove that the error magnitude of the swap operation caused by the inhomogeneous magnetic field is bounded by $\delta^2$ ($\delta = b/J$) for any initial states.

## ACKNOWLEDGEMENT

The authors thank Shu-Shen Li for useful discussions. This work is supported by the National Science Foundation of China under Grants No. 10604053.